\documentclass[12pt]{iopart}
\usepackage{xcolor}
\usepackage{graphicx}
\usepackage[colorlinks=true,linkcolor=blue,citecolor=blue,urlcolor=blue]{hyperref}

\begin{document}

\title[]{Improving quantum interference visibility between independent sources by enhancing the purity of correlated photon pairs}

\author{Hsin-Pin Lo*, Kai Asaoka \& Hiroki Takesue}

\address{Basic Research Laboratories, NTT, Inc., 3-1 Morinosato Wakamiya, Atsugi, 243-0198, Kanagawa, Japan}
\ead{hsinpin.lo@ntt.com}
\vspace{10pt}
\begin{indented}
\item[\today]
\end{indented}

\begin{abstract}
High-visibility quantum interference between independent photons is essential for demonstrating multi-photon quantum information processing, and it is closely linked to the spectral purity of correlated photon pairs. In this study, we investigate two approaches to enhance the purity of photon pairs generated from a type-0 PPLN waveguide by systematically varying both the pump bandwidth and the interference-filter bandwidth, and we directly compare their performance under identical experimental conditions.
The spectral purity is evaluated from measured joint spectral intensities using Schmidt decomposition. Both methods significantly improve the Hong–Ou–Mandel interference visibility to approximately 80$\%$. However, the former approach also yields a higher three-fold coincidence rate, which is advantageous for our ongoing efforts to increase the state fidelity and generation rate of multi-photon time-bin Greenberger–Horne–Zeilinger (GHZ) states.


\end{abstract}

%
%
%
%
%

\section{Introduction}
Quantum information processing and quantum communication are expected to play a transformative role in future technologies. Correlated photon pairs generated through spontaneous parametric down-conversion (SPDC) \cite{kwiat1995new} are widely used in many quantum technologies, such as entangled Einstein–Podolsky–Rosen (EPR) pairs \cite{PhysRev.47.777}, heralded single-photon sources \cite{Kaneda:15}, and quantum key distribution \cite{gisin2007quantum}. Quantum interference is another essential technique that enables the interaction of independent single photons and is used in protocols such as quantum teleportation \cite{PhysRevLett.70.1895,teleporationNature1997,PhysRevLett.80.1121,PhysRevLett.86.1370,Takesue:15}, entanglement swapping \cite{PhysRevLett.80.3891,PhysRevLett.88.017903,PhysRevA.71.050302,Takesue:09}, and multi-photon entanglement generation \cite{PhysRevLett.82.1345,PhysRevLett.86.4435,PhysRevLett.121.250505,Lo_2023}. To achieve high-visibility quantum interference, the interfering photons must be indistinguishable in all degrees of freedom, including polarization, temporal mode, and frequency \cite{PhysRevA.81.021801}. Therefore, the preparation of high-purity heralded photons from SPDC sources is an important topic in quantum information science \cite{horodecki2009quantum,acin2001classification,PhysRevA.98.053811,Zielnicki07062018}. Since the indistinguishability of photons is closely linked to their joint spectral structure, it is useful to describe the SPDC process in terms of its joint spectral distribution.


\begin{figure*}[t] 
\centering 
\includegraphics[width=15cm]{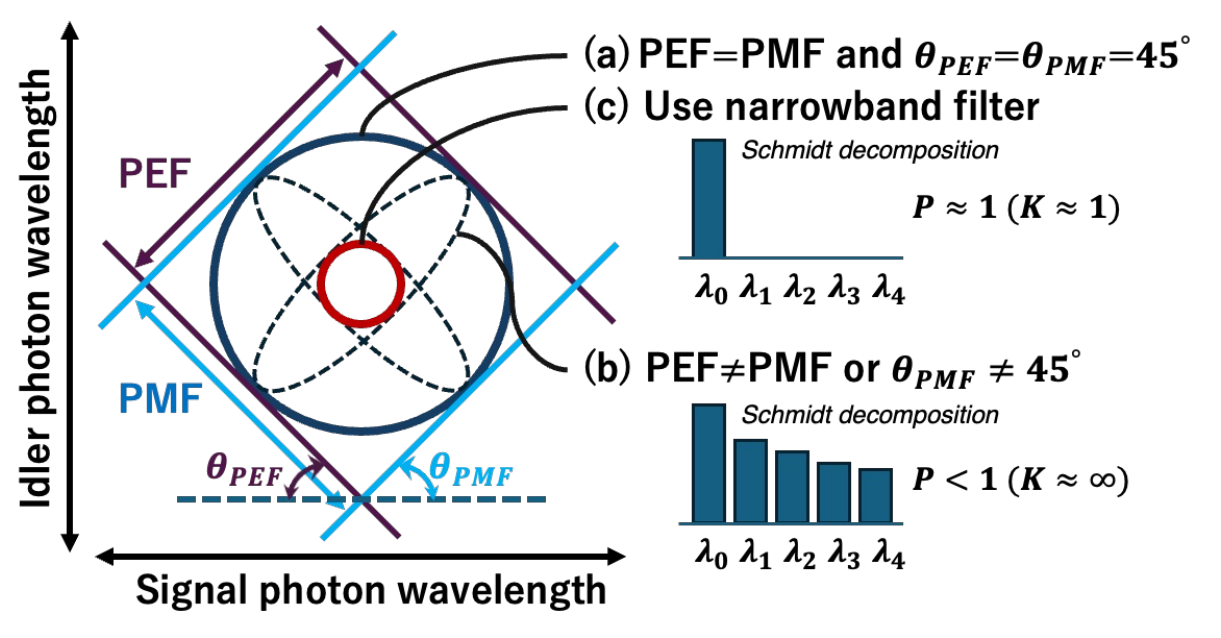}   
\caption{Schematic diagram of the joint spectrum obtained by integrating the pump envelope function (PEF) and phase matching function (PMF), where the two axes show the corresponding wavelengths of the signal photons and idle photons. (a) PEF=PMF and $\theta_{PEF}=\theta_{PMF}$, \cite{Edamatsu:2011, Jin:13} which shows a symmetric result. The Schmidt decomposition revealed high-purity SPDC photon pairs, where purity ($P$) is close to 1. (b) showed that when the PEF and PMF are not well matched, the joint spectrum is asymmetric and the purity is low. However, narrower interference filters can be used to recover high-purity photon pairs as shown in (c).} 
\label{fig1}
\end{figure*}

\begin{figure*}[t] 
\centering 
\includegraphics[width=15cm]{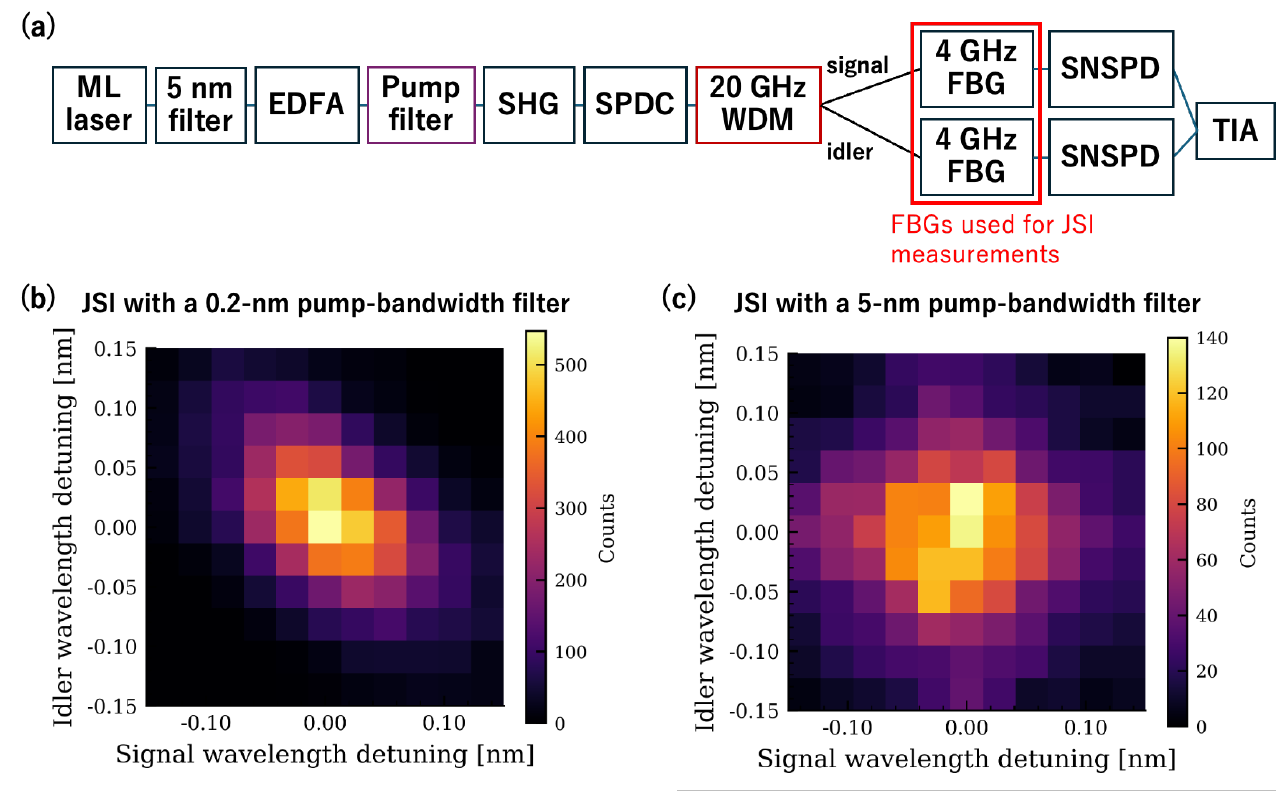}   
\caption{(a) Experimental setup. The SPDC photon pair purity was varied using a pump bandwidth tunable filter. The 4-GHz FBGs were used to perform the JSI measurements. (b) and (c) are JSI measurement results with the pump bandwidths of 0.2 nm and 5 nm, respectively. The integration time for each coincidence measurement point was 1 s. SPF: Short-wave pass edge filters. LPF: Long-wave pass edge filter. SNSPD: superconducting nanowire single-photon detector.} 
\label{fig2}
\end{figure*}

The spectral correlations in an SPDC source can be described by the two-photon state
\begin{eqnarray}
\left|\Phi \right\rangle_{s,i}=\int \int d\omega_sd\omega_iS(\omega_s,\omega_i)\hat{a}_s^\dagger(\omega_s)\hat{a}_i^\dagger(\omega_i)\left|0\right\rangle_{s,i}
\end{eqnarray}
where $\hat{a}_x^\dagger(\omega_x)$ is the creation operator of the signal (s) and idler (i) photon, respectively. $S(\omega_s,\omega_i)=\alpha(\omega_s,\omega_i)\phi(\omega_s,\omega_i)$ is the joint spectral distribution function, which depends on the spectral relation between the pump envelope function (PEF, $\alpha(\omega_s,\omega_i)$) and the phase-matching function (PMF, $\phi(\omega_s,\omega_i)$) 
as shown in the Fig.~\ref{fig1} \cite{Weston:16,PhysRevLett.84.5304,uren2006generationpurestatesinglephotonwavepackets}. The width of PMF is proportional to the inverse of the nonlinear material length. By matching the PEF and PMF, as in Fig.~\ref{fig1} (a), the symmetric joint spectrum becomes factorizable, $S(\omega_s,\omega_i)=f_s(\omega_s)f_i(\omega_i)$. The two-photon state can also be written as the sum of discrete modes
\begin{eqnarray}
\left|\Phi \right\rangle_{s,i}=\sum_n \sqrt{\lambda_n}\hat{g}_{s,n}^\dagger\hat{g}_{i,n}^\dagger\left|0\right\rangle_{s,i}, 	
\end{eqnarray}
where $\hat{g}_{x,n}=\int d\omega_x \varphi_{x,n}\hat{a}_x^\dagger(\omega_x)$, and $\varphi_{s,n}$ and $\varphi_{i,n}$ are Schmidt modes for signal and idler photons, and each mode has related Schmidt eigenvalue $\lambda_n$. The purity, $P\equiv Tr(\hat{\rho}_s^2)=\sum_n \lambda_n^2$, of the heralding signal photon from the SPDC photon pairs is defined by tracing the reduced density matrix, $\hat{\rho}_s=$Tr$_{i}$$( \left|\Phi\right\rangle_{s,i,s,i}\left\langle\Phi\right| )$. By performing the Schmidt decomposition on the factorable joint spectrum, $S(\omega_s,\omega_i)$, a highly pure heralded photon with only one spectral mode ($\lambda_0$) appears. Fig.~\ref{fig1} (b) shows the joint spectrum with lower-purity photon pairs. Narrowband interference filters \cite{Osorio_2008} can also be used to remove unwanted spectral modes ($\lambda_1, \lambda_2,...$) and recover higher purity, as shown in Fig.~\ref{fig1} (c).

Recent studies have further demonstrated the power of Schmidt mode decomposition as a versatile tool for characterizing high-dimensional photonic states and their interference properties \cite{npjQIChang2021,CHANG2025100024}.

In practice, two main approaches are commonly used to increase the spectral purity of SPDC photon pairs. One possible approach is to engineer the joint spectrum by appropriately tailoring the pump bandwidth, such that the PEF and PMF become approximately factorable \cite{PhysRevLett.105.253601,PhysRevLett.106.013603}. The other is to apply narrowband interference filters to remove undesired spectral modes \cite{Osorio_2008,PhysRevA.91.013830,SiWang2019}. The former approach can generate high-purity photons without significant loss, whereas the latter can achieve high purity at the expense of reduced brightness due to filtering loss.

Many experimental demonstrations based on type-II potassium titanyl phosphate (KTP) waveguides have shown that carefully choosing a short pump pulse duration to match the phase-matching condition enables the generation of high-purity SPDC photon pairs and high-visibility quantum interference \cite{Harder:13,Graffitti_2017,Graffitti:18}. In contrast, type-0 periodically poled lithium niobate (PPLN) waveguides, which are attractive for integrated and fiber-compatible quantum photonics, typically generate non-degenerate photon pairs with much broader spectra \cite{Fujii:07,Song:25}. In such cases, a common strategy is to employ narrowband interference filters to approximate a single spectral mode \cite{Osorio_2008,PhysRevA.91.013830,SiWang2019}; however, this inevitably reduces the photon-pair generation rate and thus limits the success probability in multi-photon experiments based on independent sources. This trade-off between spectral purity and source brightness becomes particularly critical for applications requiring multi-photon entangled states, such as time-bin Greenberger–Horne–Zeilinger (GHZ) states for quantum networking \cite{kimble2008quantum,wehner2018internet}.

In this study, we investigate how these two approaches—(i) tailoring the pump bandwidth and (ii) applying narrowband interference filters—affect both the spectral purity and the usable multi-photon generation rate in a type-0 PPLN waveguide system. We first measure the JSIs of SPDC photon pairs while varying the pump bandwidth to evaluate the resulting purities via Schmidt decomposition. We then perform Hong–Ou–Mandel (HOM) interference between independent photons using different combinations of pump bandwidths and interference filters, and compare the HOM visibility with the corresponding three-fold coincidence rates. To the best of our knowledge, this is the first direct experimental comparison of these two widely used purity-enhancement techniques under identical conditions. Our results provide practical guidelines for optimizing the balance between spectral purity and source brightness, which is essential for multi-photon quantum information processing and future scalable photonic entanglement sources.

\section{SPDC photon pair generation}
The experimental setup for preparing SPDC photon pairs is shown in Fig.~\ref{fig2} (a). Different bandwidth filters, centered at 1551.1 nm, were used to vary the pulse width of a mode-locked (ML) laser. 

As shown in Fig. 2(a), the fundamental light from the ML laser is first spectrally filtered to a bandwidth of 5 nm, amplified by an EDFA, and then further shaped by a pump-bandwidth tunable filter, which both defines the effective pump bandwidth and suppresses amplified spontaneous emission from the EDFA.

The fundamental light was inserted into a type-0 PPLN waveguide to generate second-harmonic generation (SHG) light at a central wavelength of 775.55 nm. The SHG light was then used to pump another type-0 PPLN waveguide for generating the SPDC photon pairs. The pulse durations of the fundamental and SHG lights were measured using autocorrelators and were found to be comparable after frequency doubling. An erbium-doped fiber amplifier (EDFA) was used to amplify the SHG power, and the average number of generated photon pairs per pulse was set at 0.01. It should be noted that the pump preparation scheme employing spectral filtering, EDFA amplification, and subsequent SHG is essential in our experiment. The phase-matching bandwidth of the PPLN waveguide used for SHG strongly limits the acceptable input bandwidth. Therefore, directly using a femtosecond mode-locked laser as the SHG pump would result in inefficient frequency conversion and distorted spectral profiles.

By spectrally filtering and amplifying the fundamental pulses prior to SHG, we can both ensure sufficient pump power and precisely tailor the pump bandwidth, which is critical for controlling the joint spectral properties of the generated SPDC photon pairs.

A 20-GHz-bandwidth wavelength-division multiplexer (WDM) filter was used to separate the photon pairs into signal and idler photons with central wavelengths of 1555.1 nm and 1547.1 nm, respectively.
For the JSI measurements, wavelength-tunable 4-GHz fiber Bragg grating (FBG) filters (TeraXion), providing a tuning range of $\pm$30 GHz around the central wavelength, were inserted in both the signal and idler paths.
The photons filtered by the FBGs were detected using superconducting nanowire single-photon detectors (SNSPDs), and coincidence events were recorded with a time-interval analyzer (TIA) to obtain frequency-resolved JSI data, using an integration time of 1 s per frequency bin.


\section{Joint spectrum intensity for SPDC photon pair with different pump bandwidth filters}
As shown in the previous section, the joint spectral intensity (JSI) can be used to estimate the spectral purity of photon pairs, and the effect of directly varying the pump bandwidth has already been demonstrated in earlier studies \cite{PhysRevLett.105.253601,PhysRevLett.106.013603}. Before performing quantum interference between independent photons, we first verified that the spectral purity of the SPDC photon pairs can be improved by adjusting the SHG pump pulse width, which is controlled by varying the bandwidth of the fundamental light.


Figures~\ref{fig2} (b) and (c) showed two JSI results obtained with 0.2-nm and 5-nm pump bandwidth filters, respectively. In this measurement, a 20-GHz WDM filter was used to separate the photon pairs into the signal and idler channels at 1555.1 nm and 1547.1 nm, which constrains the measured JSI. This bandwidth restriction causes the JSI to appear more symmetric when using the 5-nm pump filter, even though the intrinsic SPDC spectrum of the type-0 PPLN waveguide is highly frequency-correlated. The spectral purity was obtained using Schmidt decomposition \cite{Mosley_2008}. The estimated purities were 0.86 for the 0.2-nm pump filter and 0.98 for the 5-nm pump filter. These results confirm that tailoring the pump bandwidth through the fundamental-light filter effectively modifies the SHG pump pulse width and provides a practical means to tune the spectral purity of SPDC photon pairs.


\section{Quantum interference measurements between independent photons}

\begin{figure*}[t] 
\centering 
\includegraphics[width=15cm]{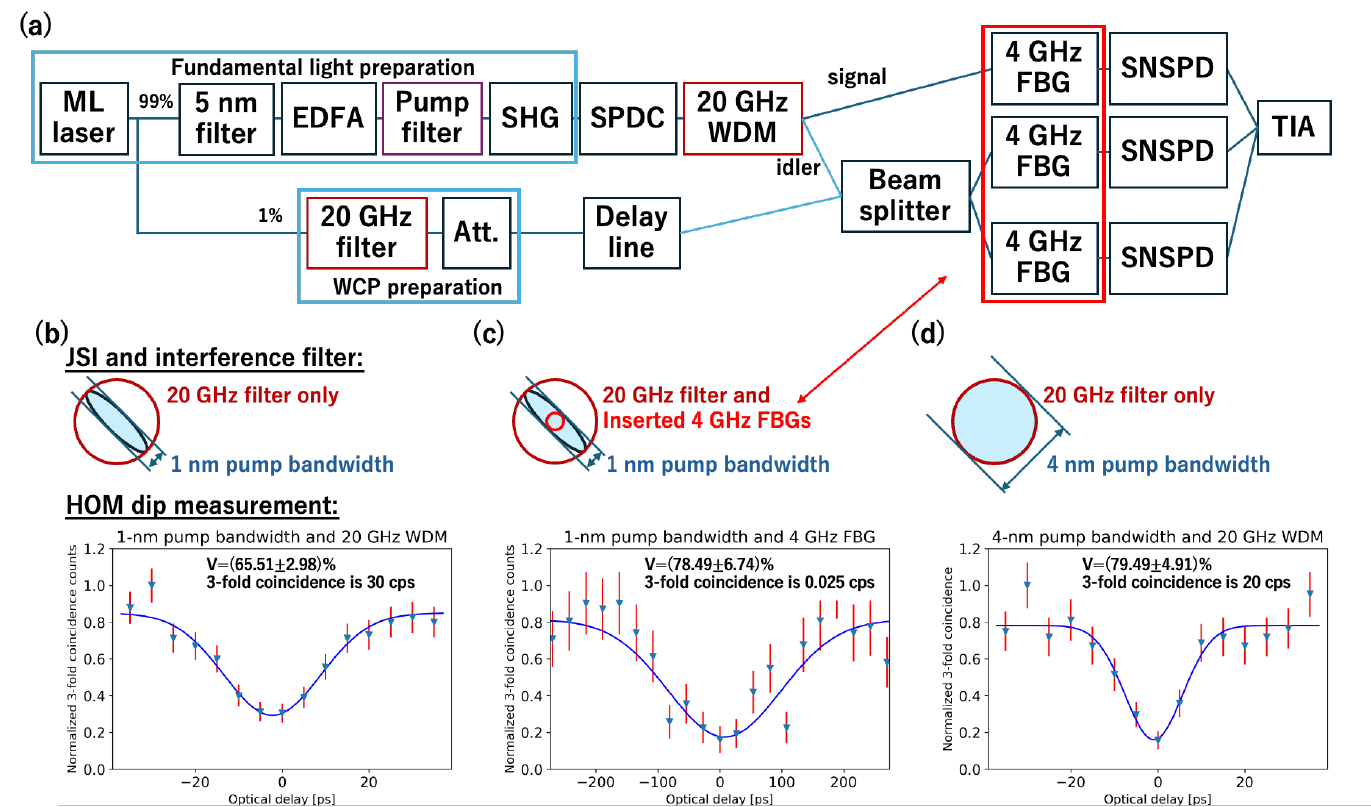}   
\caption{(a) Experimental setup for quantum interference measurement between independent light sources. (b), (c), and (d) show the HOM dip measurements performed by preparing different pump lights and different interference filter bandwidths. The upper figure shows that the JSI depends on the pump bandwidth and the interference filter bandwidth at each quantum interference measurement, and the lower figure shows the HOM dip interferometry measurement results.} 
\label{fig3}
\end{figure*}

Finally, we performed quantum interference between independent photons \cite{PhysRevLett.59.2044,Tapster01031998} using different pump bandwidths and different interference filters. The experimental setup is shown in Fig.~\ref{fig3} (a). To prepare a weak coherent pulse (WCP), 1$\%$ of the ML laser output was extracted through a fiber coupler, and a 20-GHz filter was used to generate the same central wavelength and bandwidth as the idler photons. The WCP and idler photon were then injected into a 50:50 fiber beam splitter. The temporal delay between the two photons was adjusted by fine-tuning an optical delay line to perform the Hong–Ou–Mandel (HOM) interference measurement. The average photon number per pulse of the WCP was controlled using an optical attenuator (Att.). 

It is worth clarifying that the theoretical upper limit of HOM interference visibility between a heralded single photon and a WCP remains 100$\%$, as shown in Ref. \cite{Rarity_2005,PhysRevA.83.031805}. In particular, nonclassical interference between independent sources can reach unity visibility when the heralded photon is spectrally pure and the spatial, temporal, and spectral modes are well matched.


First, we performed the quantum interference using a 1-nm pump-bandwidth filter and 20-GHz WDM filters. The HOM-dip visibility and the three-fold coincidence count outside the dip were measured to be (65.51±2.98)$\%$ and 30 counts per second (cps), respectively, as shown in Fig.~\ref{fig3} (b). The interference visibility may be limited by the spectral purity of the SPDC photon pairs, as indicated in the upper panel (“JSI and interference filter”). Therefore, we attempted to improve the purity by inserting 4-GHz FBG filters before the SNSPDs, which increased the visibility to (78.49±6.74)$\%$ but reduced the three-fold coincidence rate to 0.025 cps because of the narrowband interference filters. This result confirmed that narrow interference filters can be used to improve the purity, though the coherence time (i.e., the FWHM of the HOM dip) increases correspondingly. To achieve both high quantum interference visibility and a high three-fold coincidence rate, we used a 4-nm pump bandwidth filter and 20-GHz filters for the HOM-dip measurement, achieving (79.49±4.91)$\%$ visibility and 20 cps three-fold coincidence counts outside the dip. 

These results indicate how pump-bandwidth engineering and interference filtering contribute differently to purity, brightness, and interference visibility. However, we note that high spectral purity alone does not necessarily guarantee high HOM-interference visibility when using independent photons, since the visibility can also be degraded by factors such as spatial and temporal distinguishability, as well as photon noise. Nevertheless, the combination of high spectral purity and improved multi-photon generation rate demonstrated here provides a practical pathway toward high-state-fidelity and high-rate three-photon time-bin GHZ states \cite{Lo_2023} and entanglement-based quantum network protocols \cite{chen2008multipartitequantumcryptographicprotocols,Murta_2020,doi:10.1126/sciadv.abe0395}.

Although amplified spontaneous emission (ASE) from the EDFA is largely suppressed by the pump filter, residual pulse-shape distortion may still slightly degrade the spectral purity of the SPDC photons and consequently limit the achievable HOM visibility. Employing active spectral shaping techniques, such as programmable optical filters, could further improve the pump pulse quality.


\section{Conclusion}
The spectral purity of SPDC photon pairs is an important issue in quantum interference. We focused on achieving both high quantum-interference visibility and a high three-photon pair generation rate by using different pump-bandwidth filters and different bandwidths of interference filters. In this study, we confirmed that the spectral purity of SPDC photon pairs can be tuned by adjusting the pulse width of the fundamental light. From the HOM-dip measurements between independent photons, we found that using a 4-nm pump-bandwidth filter improves the visibility to around 80$\%$ while maintaining a much higher three-photon generation rate compared with narrowband interference filtering. Together with the comparison between pump-bandwidth engineering and narrowband interference filtering performed in this work, these results provide practical guidelines for optimizing the trade-off between purity and brightness. We believe that this research will support the realization of high-fidelity, high-rate multi-photon time-bin GHZ states and entanglement-based quantum networks in subsequent studies.


\section*{Acknowledgement}
This work was partially supported by MIC R$\&$D of ICT Priority Technology (JPMI00316).

\section*{References}
\bibliographystyle{iopart-num}
\bibliography{JJAP_resubmit_ref}

\providecommand{\newblock}{}
\begin{thebibliography}{10}
\expandafter\ifx\csname url\endcsname\relax
  \def\url#1{{\tt #1}}\fi
\expandafter\ifx\csname urlprefix\endcsname\relax\def\urlprefix{URL }\fi
\providecommand{\eprint}[2][]{\url{#2}}

\bibitem{kwiat1995new}
Kwiat P~G, Mattle K, Weinfurter H, Zeilinger A, Sergienko A~V and Shih Y 1995
  {\em Physical Review Letters\/} {\bf 75} 4337

\bibitem{PhysRev.47.777}
Einstein A, Podolsky B and Rosen N 1935 {\em Phys. Rev.\/} {\bf 47}(10)
  777--780 \urlprefix\url{https://link.aps.org/doi/10.1103/PhysRev.47.777}

\bibitem{Kaneda:15}
Kaneda F, Christensen B~G, Wong J~J, Park H~S, McCusker K~T and Kwiat P~G 2015
  {\em Optica\/} {\bf 2} 1010--1013
  \urlprefix\url{https://opg.optica.org/optica/abstract.cfm?URI=optica-2-12-1010}

\bibitem{gisin2007quantum}
Gisin N and Thew R 2007 {\em Nature Photonics\/} {\bf 1} 165--171

\bibitem{PhysRevLett.70.1895}
Bennett C~H, Brassard G, Cr\'epeau C, Jozsa R, Peres A and Wootters W~K 1993
  {\em Phys. Rev. Lett.\/} {\bf 70}(13) 1895--1899
  \urlprefix\url{https://link.aps.org/doi/10.1103/PhysRevLett.70.1895}

\bibitem{teleporationNature1997}
D~Bouwmeester J-W~Pan K~M~M~E~H~W and Zeilinger A 1997 {\em Nature\/} {\bf
  390}(6660) 575--579

\bibitem{PhysRevLett.80.1121}
Boschi D, Branca S, De~Martini F, Hardy L and Popescu S 1998 {\em Phys. Rev.
  Lett.\/} {\bf 80}(6) 1121--1125
  \urlprefix\url{https://link.aps.org/doi/10.1103/PhysRevLett.80.1121}

\bibitem{PhysRevLett.86.1370}
Kim Y~H, Kulik S~P and Shih Y 2001 {\em Phys. Rev. Lett.\/} {\bf 86}(7)
  1370--1373
  \urlprefix\url{https://link.aps.org/doi/10.1103/PhysRevLett.86.1370}

\bibitem{Takesue:15}
Takesue H, Dyer S~D, Stevens M~J, Verma V, Mirin R~P and Nam S~W 2015 {\em
  Optica\/} {\bf 2} 832--835
  \urlprefix\url{https://opg.optica.org/optica/abstract.cfm?URI=optica-2-10-832}

\bibitem{PhysRevLett.80.3891}
Pan J~W, Bouwmeester D, Weinfurter H and Zeilinger A 1998 {\em Phys. Rev.
  Lett.\/} {\bf 80}(18) 3891--3894
  \urlprefix\url{https://link.aps.org/doi/10.1103/PhysRevLett.80.3891}

\bibitem{PhysRevLett.88.017903}
Jennewein T, Weihs G, Pan J~W and Zeilinger A 2001 {\em Phys. Rev. Lett.\/}
  {\bf 88}(1) 017903
  \urlprefix\url{https://link.aps.org/doi/10.1103/PhysRevLett.88.017903}

\bibitem{PhysRevA.71.050302}
de~Riedmatten H, Marcikic I, van Houwelingen J~A~W, Tittel W, Zbinden H and
  Gisin N 2005 {\em Phys. Rev. A\/} {\bf 71}(5) 050302
  \urlprefix\url{https://link.aps.org/doi/10.1103/PhysRevA.71.050302}

\bibitem{Takesue:09}
Takesue H and Miquel B 2009 {\em Opt. Express\/} {\bf 17} 10748--10756
  \urlprefix\url{https://opg.optica.org/oe/abstract.cfm?URI=oe-17-13-10748}

\bibitem{PhysRevLett.82.1345}
Bouwmeester D, Pan J~W, Daniell M, Weinfurter H and Zeilinger A 1999 {\em Phys.
  Rev. Lett.\/} {\bf 82}(7) 1345--1349
  \urlprefix\url{https://link.aps.org/doi/10.1103/PhysRevLett.82.1345}

\bibitem{PhysRevLett.86.4435}
Pan J~W, Daniell M, Gasparoni S, Weihs G and Zeilinger A 2001 {\em Phys. Rev.
  Lett.\/} {\bf 86}(20) 4435--4438
  \urlprefix\url{https://link.aps.org/doi/10.1103/PhysRevLett.86.4435}

\bibitem{PhysRevLett.121.250505}
Zhong H~S, Li Y, Li W, Peng L~C, Su Z~E, Hu Y, He Y~M, Ding X, Zhang W, Li H,
  Zhang L, Wang Z, You L, Wang X~L, Jiang X, Li L, Chen Y~A, Liu N~L, Lu C~Y
  and Pan J~W 2018 {\em Phys. Rev. Lett.\/} {\bf 121}(25) 250505
  \urlprefix\url{https://link.aps.org/doi/10.1103/PhysRevLett.121.250505}

\bibitem{Lo_2023}
Lo H~P, Ikuta T, Azuma K, Honjo T, Munro W~J and Takesue H 2023 {\em Quantum
  Science and Technology\/} {\bf 8} 035003
  \urlprefix\url{https://dx.doi.org/10.1088/2058-9565/acc7c2}

\bibitem{PhysRevA.81.021801}
Aboussouan P, Alibart O, Ostrowsky D~B, Baldi P and Tanzilli S 2010 {\em Phys.
  Rev. A\/} {\bf 81}(2) 021801
  \urlprefix\url{https://link.aps.org/doi/10.1103/PhysRevA.81.021801}

\bibitem{horodecki2009quantum}
Horodecki R, Horodecki P, Horodecki M and Horodecki K 2009 {\em Reviews of
  Modern Physics\/} {\bf 81} 865

\bibitem{acin2001classification}
Ac{\'\i}n A, Bru{\ss} D, Lewenstein M and Sanpera A 2001 {\em Physical Review
  Letters\/} {\bf 87} 040401

\bibitem{PhysRevA.98.053811}
Graffitti F, Kelly-Massicotte J, Fedrizzi A and Bra\ifmmode~\acute{n}\else
  \'{n}\fi{}czyk A~M 2018 {\em Phys. Rev. A\/} {\bf 98}(5) 053811
  \urlprefix\url{https://link.aps.org/doi/10.1103/PhysRevA.98.053811}

\bibitem{Zielnicki07062018}
Zielnicki K, Garay-Palmett K, Cruz-Delgado D, Cruz-Ramirez H, O’Boyle M~F,
  Fang B, Lorenz V~O, U’Ren A~B and Kwiat P~G 2018 {\em Journal of Modern
  Optics\/} {\bf 65} 1141--1160 (\textit{Preprint}
  \eprint{https://doi.org/10.1080/09500340.2018.1437228})
  \urlprefix\url{https://doi.org/10.1080/09500340.2018.1437228}

\bibitem{Edamatsu:2011}
Keiichi~Edamatsu Ryosuke~Shimizu W~U~R~B~J~F~K~M~Y~H~S~S~N~A~S and Suizu K 2011
  {\em Progress in Informatics\/}  19--26
  \urlprefix\url{https://www.nii.ac.jp/pi/n8/8_19.html}

\bibitem{Jin:13}
Jin R~B, Shimizu R, Wakui K, Benichi H and Sasaki M 2013 {\em Opt. Express\/}
  {\bf 21} 10659--10666
  \urlprefix\url{https://opg.optica.org/oe/abstract.cfm?URI=oe-21-9-10659}

\bibitem{Weston:16}
Weston M~M, Chrzanowski H~M, Wollmann S, Boston A, Ho J, Shalm L~K, Verma V~B,
  Allman M~S, Nam S~W, Patel R~B, Slussarenko S and Pryde G~J 2016 {\em Opt.
  Express\/} {\bf 24} 10869--10879
  \urlprefix\url{https://opg.optica.org/oe/abstract.cfm?URI=oe-24-10-10869}

\bibitem{PhysRevLett.84.5304}
Law C~K, Walmsley I~A and Eberly J~H 2000 {\em Phys. Rev. Lett.\/} {\bf 84}(23)
  5304--5307
  \urlprefix\url{https://link.aps.org/doi/10.1103/PhysRevLett.84.5304}

\bibitem{uren2006generationpurestatesinglephotonwavepackets}
U'Ren A~B, Silberhorn C, Erdmann R, Banaszek K, Grice W~P, Walmsley I~A and
  Raymer M~G 2006 Generation of pure-state single-photon wavepackets by
  conditional preparation based on spontaneous parametric downconversion
  (\textit{Preprint} \eprint{quant-ph/0611019})
  \urlprefix\url{https://arxiv.org/abs/quant-ph/0611019}

\bibitem{Osorio_2008}
Osorio C~I, Valencia A and Torres J~P 2008 {\em New Journal of Physics\/} {\bf
  10} 113012 \urlprefix\url{https://dx.doi.org/10.1088/1367-2630/10/11/113012}

\bibitem{npjQIChang2021}
Kai-Chi~Chang Xiang~Cheng M~C~S~A~K~V~Y~S~L~T~Z~Y~X~G~Z~X~J~H~S~F~N~C~W and
  Wong C~W 2021 {\em npj Quantum Information\/} {\bf 7}(48) 1--11

\bibitem{CHANG2025100024}
Chang K~C, Cheng X, Sarihan M~C and Wong C~W 2025 {\em Newton\/} {\bf 1} 100024
  ISSN 2950-6360
  \urlprefix\url{https://www.sciencedirect.com/science/article/pii/S2950636025000167}

\bibitem{PhysRevLett.105.253601}
Evans P~G, Bennink R~S, Grice W~P, Humble T~S and Schaake J 2010 {\em Phys.
  Rev. Lett.\/} {\bf 105}(25) 253601
  \urlprefix\url{https://link.aps.org/doi/10.1103/PhysRevLett.105.253601}

\bibitem{PhysRevLett.106.013603}
Eckstein A, Christ A, Mosley P~J and Silberhorn C 2011 {\em Phys. Rev. Lett.\/}
  {\bf 106}(1) 013603
  \urlprefix\url{https://link.aps.org/doi/10.1103/PhysRevLett.106.013603}

\bibitem{PhysRevA.91.013830}
Gerrits T, Marsili F, Verma V~B, Shalm L~K, Shaw M, Mirin R~P and Nam S~W 2015
  {\em Phys. Rev. A\/} {\bf 91}(1) 013830
  \urlprefix\url{https://link.aps.org/doi/10.1103/PhysRevA.91.013830}

\bibitem{SiWang2019}
Si~Wang Chen-Xi~Liu J~L and Wang Q 2019 {\em Scientific reports\/} {\bf 9} 1--7
  (\textit{Preprint}
  \eprint{https://www.nature.com/articles/s41598-019-40720-5})
  \urlprefix\url{https://www.nature.com/articles/s41598-019-40720-5}

\bibitem{Harder:13}
Harder G, Ansari V, Brecht B, Dirmeier T, Marquardt C and Silberhorn C 2013
  {\em Opt. Express\/} {\bf 21} 13975--13985
  \urlprefix\url{https://opg.optica.org/oe/abstract.cfm?URI=oe-21-12-13975}

\bibitem{Graffitti_2017}
Graffitti F, Kundys D, Reid D~T, Brańczyk A~M and Fedrizzi A 2017 {\em Quantum
  Science and Technology\/} {\bf 2} 035001
  \urlprefix\url{https://dx.doi.org/10.1088/2058-9565/aa78d4}

\bibitem{Graffitti:18}
Graffitti F, Barrow P, Proietti M, Kundys D and Fedrizzi A 2018 {\em Optica\/}
  {\bf 5} 514--517
  \urlprefix\url{https://opg.optica.org/optica/abstract.cfm?URI=optica-5-5-514}

\bibitem{Fujii:07}
Fujii G, Namekata N, Motoya M, Kurimura S and Inoue S 2007 {\em Opt. Express\/}
  {\bf 15} 12769--12776
  \urlprefix\url{https://opg.optica.org/oe/abstract.cfm?URI=oe-15-20-12769}

\bibitem{Song:25}
Song D, Gao L, Wang D, Jin R, Jin Y, Yuan C, Cai H, Wang S, Luo Q, Wang J and
  Wang Y 2025 {\em Opt. Continuum\/} {\bf 4} 1380--1394
  \urlprefix\url{https://opg.optica.org/optcon/abstract.cfm?URI=optcon-4-7-1380}

\bibitem{kimble2008quantum}
Kimble H~J 2018 {\em Nature\/} {\bf 362} 1--10

\bibitem{wehner2018internet}
Stephanie~Wehner D~E and Hanson R 2008 {\em Science\/} {\bf 453} 1023--1030

\bibitem{Mosley_2008}
Mosley P~J, Lundeen J~S, Smith B~J and Walmsley I~A 2008 {\em New Journal of
  Physics\/} {\bf 10} 093011
  \urlprefix\url{https://dx.doi.org/10.1088/1367-2630/10/9/093011}

\bibitem{PhysRevLett.59.2044}
Hong C~K, Ou Z~Y and Mandel L 1987 {\em Phys. Rev. Lett.\/} {\bf 59}(18)
  2044--2046
  \urlprefix\url{https://link.aps.org/doi/10.1103/PhysRevLett.59.2044}

\bibitem{Tapster01031998}
Tapster P~R and and J~G~R 1998 {\em Journal of Modern Optics\/} {\bf 45}
  595--604 (\textit{Preprint}
  \eprint{https://doi.org/10.1080/09500349808231917})
  \urlprefix\url{https://doi.org/10.1080/09500349808231917}

\bibitem{Rarity_2005}
Rarity J~G, Tapster P~R and Loudon R 2005 {\em Journal of Optics B: Quantum and
  Semiclassical Optics\/} {\bf 7} S171
  \urlprefix\url{https://doi.org/10.1088/1464-4266/7/7/007}

\bibitem{PhysRevA.83.031805}
Jin R~B, Zhang J, Shimizu R, Matsuda N, Mitsumori Y, Kosaka H and Edamatsu K
  2011 {\em Phys. Rev. A\/} {\bf 83}(3) 031805
  \urlprefix\url{https://link.aps.org/doi/10.1103/PhysRevA.83.031805}

\bibitem{chen2008multipartitequantumcryptographicprotocols}
Chen K and Lo H~K 2008 Multi-partite quantum cryptographic protocols with noisy
  ghz states (\textit{Preprint} \eprint{quant-ph/0404133})
  \urlprefix\url{https://arxiv.org/abs/quant-ph/0404133}

\bibitem{Murta_2020}
Murta G, Grasselli F, Kampermann H and Bruß D 2020 {\em Advanced Quantum
  Technologies\/} {\bf 3} 2000025 (\textit{Preprint}
  \eprint{https://advanced.onlinelibrary.wiley.com/doi/pdf/10.1002/qute.202000025})
  \urlprefix\url{https://advanced.onlinelibrary.wiley.com/doi/abs/10.1002/qute.202000025}

\bibitem{doi:10.1126/sciadv.abe0395}
Proietti M, Ho J, Grasselli F, Barrow P, Malik M and Fedrizzi A 2021 {\em
  Science Advances\/} {\bf 7} eabe0395 (\textit{Preprint}
  \eprint{https://www.science.org/doi/pdf/10.1126/sciadv.abe0395})
  \urlprefix\url{https://www.science.org/doi/abs/10.1126/sciadv.abe0395}

\end{thebibliography}

\end{document}